# Quantifying Defects in Thin Films using Machine Vision

N. Taherimakhsousi[1], B. P. MacLeod[1,2], F. G. L. Parlane[1,2], T. D. Morrissey[1,2], E. P. Booker[1], K. E. Dettelbach[1], C. P. Berlinguette[1-4]*

**Affiliations**

[1]Department of Chemistry, The University of British Columbia, Vancouver, British Columbia, Canada

[2]Stewart Blusson Quantum Matter Institute, The University of British Columbia, Vancouver, British Columbia, Canada

[3]Department of Chemical & Biological Engineering, The University of British Columbia, Vancouver, British Columbia, Canada

[4]Canadian Institute for Advanced Research (CIFAR), MaRS Centre, Toronto, Ontario, Canada

*Email: cberling@chem.ubc.ca

## Abstract

The sensitivity of thin-film materials and devices to defects motivates extensive research into the optimization of film morphology. This research could be accelerated by automated experiments that characterize the response of film morphology to synthesis conditions. Optical imaging can resolve morphological defects in thin films and is readily integrated into automated experiments but the large volumes of images produced by such systems require automated analysis. Existing approaches to automatically analyzing film morphologies in optical images require application-specific customization by software experts and are not robust to changes in image content or imaging conditions. Here we present a versatile convolutional neural network (CNN) for thin-film image analysis which can identify and quantify the extent of a variety of defects and is applicable to multiple materials and imaging conditions. This CNN is readily adapted to new thin-film image analysis tasks and will facilitate the use of imaging in automated thin-film research systems.

## Introduction

Film morphology optimization is important for reducing the detrimental impacts of defects on the performance of thin-film devices such as photovoltaics[1,2] and light-emitting diodes[3]. Images of thin films carry information about common morphological defects, such as cracking[4] and dewetting[5,6], which are controlled by film synthesis conditions.While automated experiments can generate images of thin films synthesized under numerous distinct conditions, existing approaches to automatically analyzing film morphologies in such images typically require application-specific customization by software experts and are not robust to changes in image content or imaging conditions[7,8]. Here we present a versatile convolutional neural network (CNN) for thin-film image analysis which can identify and quantify the extent of a variety of defects, is applicable to multiple materials and imaging conditions and is readily adapted to new thin-film image analysis tasks.

The severity of film defects such as thickness variations, cracks, precipitates, or dewetting can often be identified by the naked eye or with optical microscopy[4,9,10]. For this reason, rapid, non-destructive, optical inspection of thin films is often carried out in the place of more expensive, more destructive, or more time-consuming methods such as stylus profilometry, atomic force microscopy, or electron microscopy. Quantitative defect analysis enables researchers to identify potentially subtle trends in film morphology as a function of experimental conditions. Researchers frequently perform quantitative image analyses using semi-manual software tools, such as measuring film coverage with *ImageJ*[11], or surface roughness with *Gwyddion*[12]. Semi-manual analysis becomes impractical, however, when applied to high-throughput experiments or high-speed manufacturing where images of thin films are generated at high frequency or in large numbers. In such cases, automated image analysis is necessary. Automated analyses of images of thin-film materials and devices are often performed using image-processing algorithms which are specific to the material, morphology, and imaging modality of interest[13]. An example of this type of approach is the matrix-based analysis of orientational order in AFM images of P3HT nanofibers[7]. This traditional type of computer vision includes application-specific feature extraction subroutines with numerous adjustable parameters, such as imaging condition-dependant thresholds, which can make them difficult to adapt to new applications[14], such as new materials, morphologies, or imaging modalities. Here, we describe a new approach to image-based thin-film defect quantification which uses a CNN to overcome many limitations of previous approaches. The CNN we developed for this purpose, which we call *DeepThin,* quantifies the extent of several types of common morphological defects (e.g., particles, cracks, scratches, and dewetting) in images of thin films. We show that *DeepThin* works with different imaging modalities (dark- and bright-field microscopy), different magnifications and different materials (a small-molecule organic glass and a metal oxide) and can readily be retrained to detect new defect types.

CNNs are a family of machine learning algorithms that have been applied to image classification[15], feature detection[16], image segmentation[17], and object recognition problems[18]. CNNs have achieved classification accuracy comparable to human experts in computer vision challenges[19,20]. The performance and robustness demonstrated by CNNs makes them appealing for thin-film defect analysis. An additional benefit of CNNs is that they can be easily trained using examples provided by a domain expert (e.g. a materials scientist) rather than through involved algorithm customization by a computer vision expert[14]. CNNs are an established approach to electron microscopy image analysis tasks, with examples in the materials sciences including mechanical property estimation[21], nanoparticle segmentation[22], and nanostructure classification[23]. However, CNNs have only recently been applied for the analysis of optical images of thin films in two highly-application-specific ways: classifying the corrosion conditions under which surface films formed on metal surfaces[24,25] and determining the thickness of exfoliated 2D crystals[26]. To the best of our knowledge, the *DeepThin* CNN reported here is the first example of a general-purpose CNN for classifying or quantifying common morphological defects in optical images of thin films.

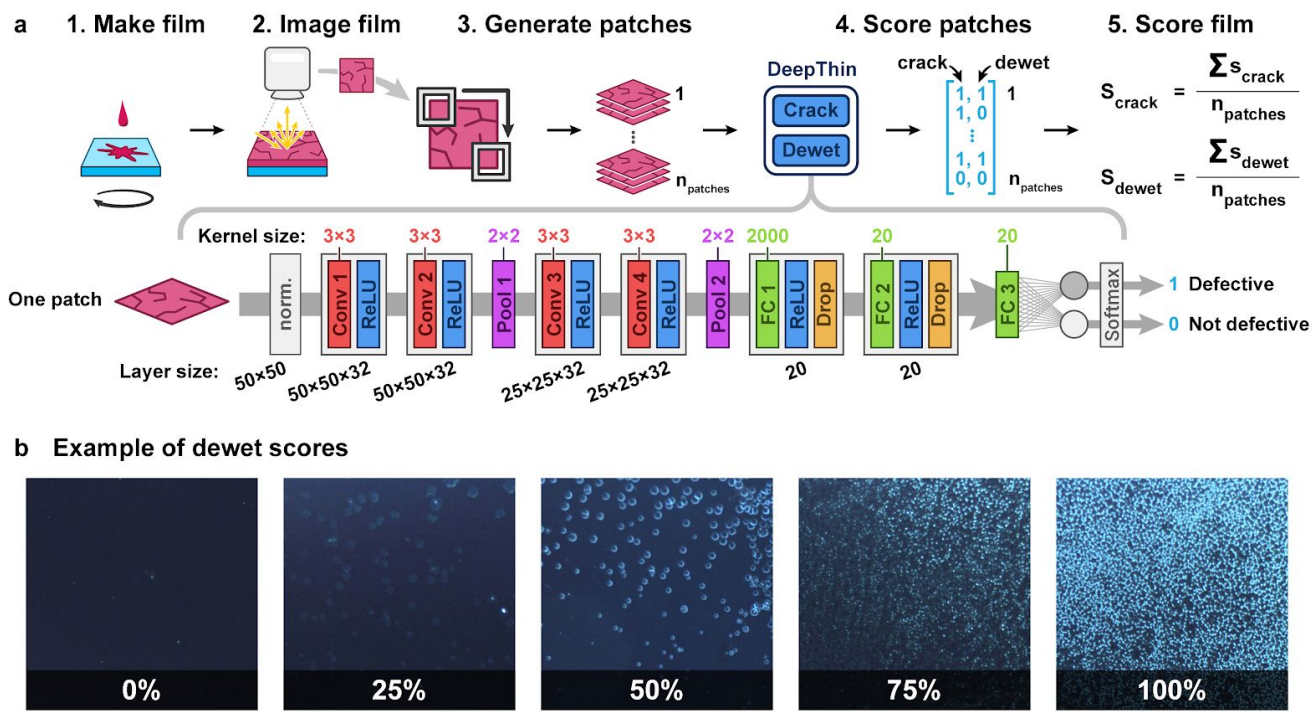

**Fig. 1 | Using *DeepThin* to evaluate the morphology of a thin film. a**, Scheme for applying *DeepThin* to score the extent of cracking ($S_{crack}$) and dewetting ($S_{dewet}$): (1) Thin film sample is created, for example by spin-coating. (2) A darkfield photograph of the sample is taken. (3) The image is subdivided into n patches (4) The *DeepThin* convolutional neural network (architecture shown) is applied to each patch (5) Scores for the extent of cracking and dewetting are computed. The structure of *DeepThin* is shown (bottom). The numbers below each layer indicate the output data size of each convolution or fully connected layer. Conv: Convolution layer; Pool: Pooling layer; ReLU: Rectified Linear Units layer; FC: fully connected layer. **b**, Example images of organic thin films from the testing portion of the darkfield dataset with varying extents of cracking and dewetting, ordered by the scores assigned to them by *DeepThin*.

## Results

### Training dataset and model development

To develop and validate *DeepThin* (Fig. 1), we first created a dataset of 2600 darkfield images of organic semiconductor thin films (each 4000×3000 pixels) exhibiting varying extents of cracking and dewetting due to differences in film composition and annealing conditions. These films were deposited by spin-coating, annealed, and imaged by a flexible robotic platform equipped with a darkfield photography system (Methods, ref.[27]). The images in this darkfield dataset were labelled with respect to the extent of dewetting and of cracking by materials scientists with expertise in thin-film materials research. Labelling was on a subjective integer scale from zero (no defects observed) to ten (extremely defected) for both defect types. The dataset was augmented by applying rotations and mirroring to the labelled images to obtain a total of 17,374 labelled images. This dataset was then randomly divided into training, validation and test sets as detailed in Table S1 to facilitate the development of a CNN for image-based thin film defect analysis. We evaluated the suitability of several state-of-the-art CNNs architectures[28,29] for this task before choosing to develop a new architecture for *DeepThin* (Methods) inspired by the VGG16 CNN (Methods). We then optimized the weights and biases of *DeepThin* using the *Adam* optimizer (Methods). After this optimization, *DeepThin* scored images of cracked and dewetted films in the darkfield dataset with >93% accuracy (Table S2).

## *Model validation against a known, monotonic morphological trend*

To further validate *DeepThin* we carried out an experiment where an organic semiconductor film was imaged as it underwent thermally-activated dewetting[9] (Fig. 2a). This experiment provided a series of images in which the extent of dewetting was known to increase monotonically with respect to time. We then used *DeepThin* to quantify the extent of dewetting in each image. The resulting dewetting scores also increased monotonically with respect to time (Fig. 2b), showing that *DeepThin* can correctly order a set of images of thin films based on a one-dimensional trend in the film morphology.

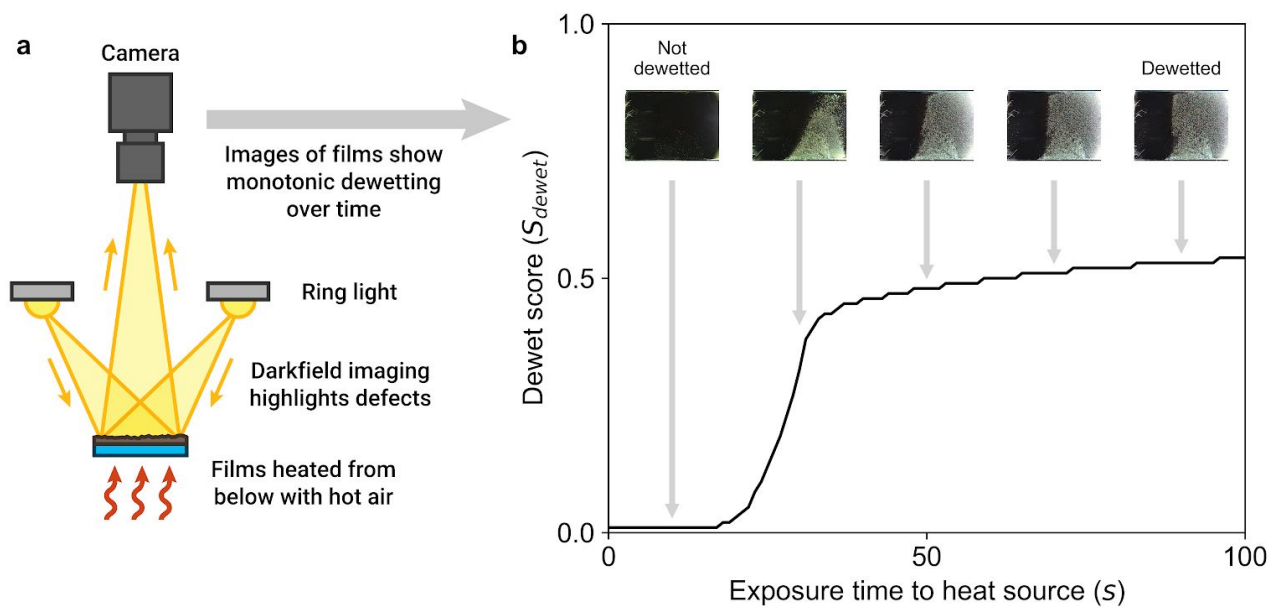

**Fig. 2 | Recovery of a monotonic trend in film morphology using *DeepThin*. a**, The experimental setup used for capturing a series of images of a thin film with an extent of dewetting which increases monotonically in time. **b**, Dewetting score assigned to images of the thin film sample as a function of heating time. The extent of dewetting increased monotonically with time throughout the experiment as seen in the images from 10, 30, 50, 70, and 90 seconds into the experiment. Similarly, the dewetting score reported by *DeepThin* also increases monotonically.

## *Resolution of a two-dimensional film-morphology response surface*

To illustrate the applicability of our method to thin film optimization, we used *DeepThin* to resolve a 2-dimensional film-morphology response surface in a set of experiments where both film composition and processing were varied. Following our previous work[27], thin films of spiro-OMeTAD doped with varying amounts of FK102 Co(III) TFSI and annealed for varying durations were prepared and then imaged using a robotic platform (Methods). These experiments provided an array of images exhibiting morphological trends as a function of both film composition and processing. The analysis of these images using *DeepThin* automatically provided a response surface quantifying the extent of dewetting as a function of the film composition and annealing time (Fig. 3). From this surface, two trends can readily be identified:(i) the extent of dewetting increased as the dopant-to-spiro-OMeTAD molar ratio increased from 0 to 0.4, then decreased at higher dopant levels; (ii) longer annealing times produced more dewetted films across all dopant-to-spiro-OMeTAD ratios, with the exception of undoped films which did not exhibit dewetting regardless of annealing time. The ability to automatically obtain composition-processing-morphology response surfaces such as the one shown in Fig. 3 using rapid, inexpensive and non-destructive imaging is a benefit of our approach.

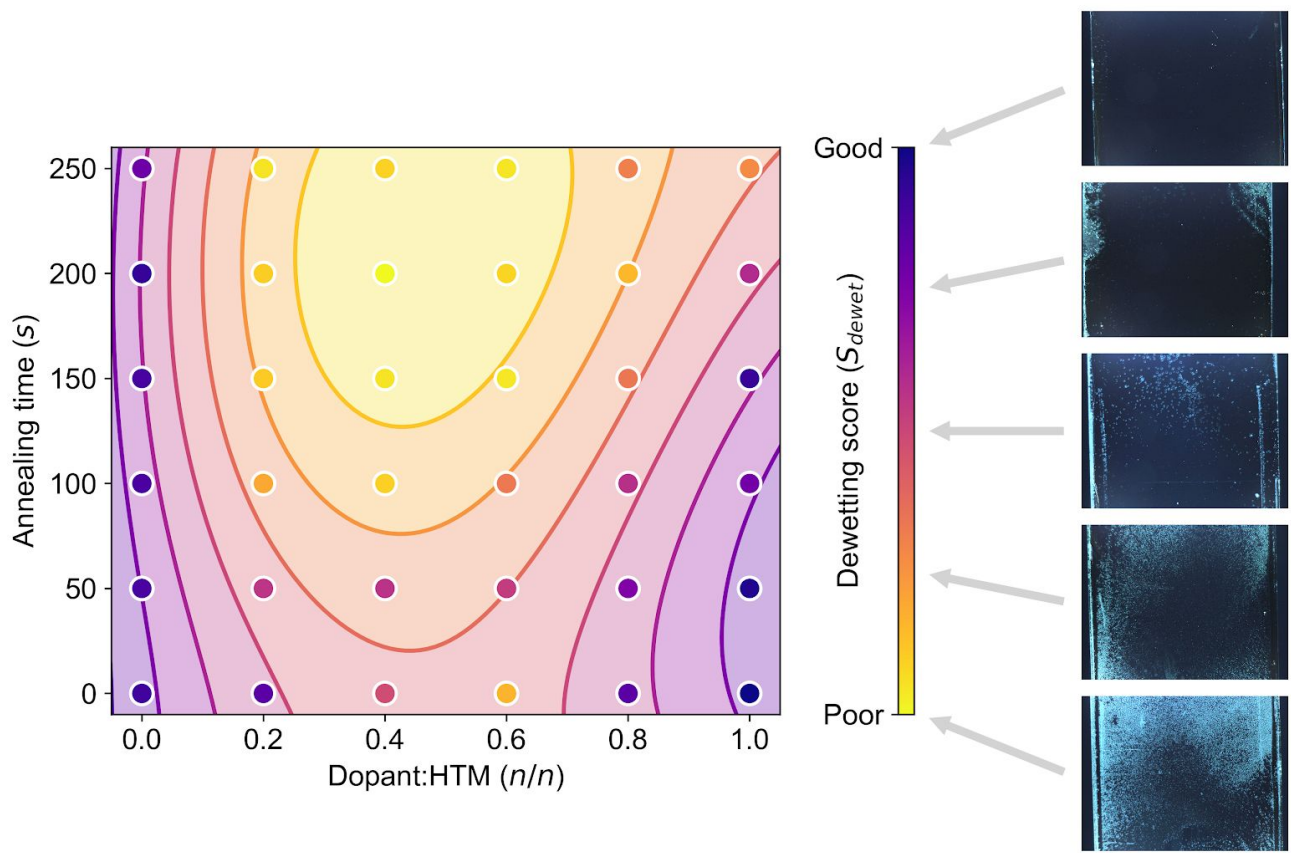

**Fig. 3 | Use of *DeepThin* to resolve trends in morphology caused by variations in composition and processing of organic thin films.** The thermally-activated dewetting of the organic semiconductor film was suppressed for levels of p-doping below 0.2 or above 0.8 whereas at intermediate doping levels, dewetting occurred after annealing above 50 s.

*Applicability of DeepThin to multiple materials, defect types and imaging modalities*

To demonstrate the versatility of *DeepThin*, we next applied it to a different imaging modality (bright-field microscopy) at different magnifications (5× and 20×), to additional defect types (scratches, particles, and thickness non-uniformities) and to films of a different material (a metal oxide) (Fig. 4). For these demonstrations, three new image datasets were manually obtained using a bright-field microscope (Methods): a set of 129 images of organic semiconductor films at 5× magnification and two sets of images of $TiO_x$ films (81 images at 5× magnification and 82 at 20× magnification). These microscope images, originally 1024×768 pixels, were divided into 100×100 pixels patches, manually labelled based on the types of defects present and then subjected to reflections and rotations to obtain augmented data sets of adequate size for retraining and testing *DeepThin* (Tables S3-S5). These images were labelled for cracking, dewetting and for additional defect types not considered in the darkfield dataset originally used for model development (scratches, particles, and thickness non-uniformities). After a separate retraining for each of the three microscopy data sets, *DeepThin* was able to accurately detect the five labelled defect types (cracks, dewetting, particles, scratches and thickness non-uniformities) wherever they appeared in the datasets for the different materials and magnifications (Fig. 4, Tables S6, S7). These results demonstrate that CNNs such as *DeepThin* may be applied to a broad scope of thin-film materials, defect morphologies and imaging conditions.

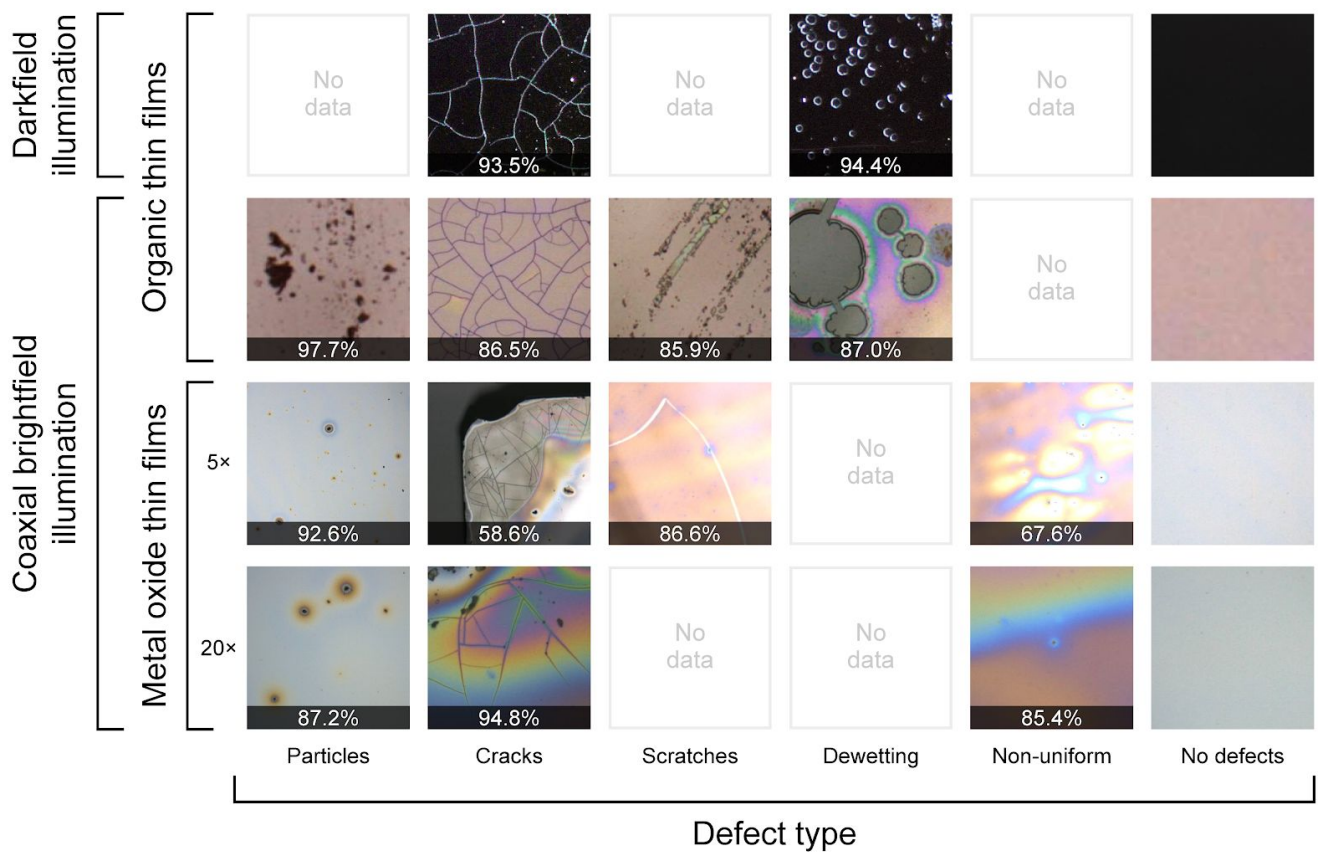

**Fig. 4 | Accuracy of *DeepThin* under a variety of conditions.** In the top row, the accuracy for reproducing the human-labeled scores for the extent of cracking and extent of dewetting in an image is shown. In all other rows, the accuracy for correctly classifying the images based on the presence or absence of different morphological defects is shown. Empty cells in the figure are associated with defects that were not present in the datasets or were not labelled for this study.

### *Benchmarking against concrete defect detection literature*

To assess the performance of *DeepThin* for defect detection in a domain other than thin films, we benchmarked *DeepThin* against previously reported state-of-the-art algorithms for crack detection[30] and segmentation[31,32] in images of concrete and road surfaces. We used a dataset provided by Zhang and coworkers[30] to benchmark the road-surface crack detection ability of *DeepThin*. The accuracy statistics given in Table S9 and the receiver operating characteristic (ROC) curves in Fig. S1 show that *DeepThin* outperforms the three crack detection algorithms used by Zhang and coworkers.[30] Next, we benchmarked *DeepThin* against the road surface crack segmentation algorithms described by Chen et al.[31] and Li et al.[32] using the 118 image CFD dataset provided by Chen et al. We used 72 of these images for training and 46 images for testing as was done by Chen et al. and again found that *DeepThin* again achieves state-of-art performance (Tables S11 and S12). These results suggest that *DeepThin* may also have utility in areas of materials science other than thin films.

## Discussion

We have shown that a convolutional neural network can accurately identify several different types of morphological defects in images of organic and inorganic thin films acquired under a variety of imaging conditions. The versatility of this approach to defect detection arises due to the ease with which it can be adapted to new defect types using labelled example images. The labelling of images containing examples of materials defects provides a straightforward mechanism for materials scientists to encode their domain expertise into an image analysis algorithm. As this example-based process for algorithm customization does not require software engineering expertise, we expect CNN-based approaches to material defect analysis to increase the accessibility of automated image analysis to the materials science community. Our *DeepThin* CNN provides the ability to rapidly and automatically identify trends in film morphology arising from manipulations of composition and process variables. We anticipate that capabilities of this kind, particularly in combination with automated experimentation, will accelerate thin film materials science research by facilitating the optimization of materials in design spaces where the morphological response to the experimental parameters is initially unknown.

## Methods

### *Robotic platform for film deposition, annealing, and imaging*

Deposition, annealing and darkfield imaging of all the organic thin films included in the database were performed using a flexible robotic platform configured for thin-film experiments described in detail in ref.[27] Briefly, the robotic platform consists of a multi-purpose robotic arm that can handle fluids and planar glass substrates, as well as a variety of other modules which enable other tasks to be performed. The modules relevant to this study include: trays of stock solutions and mixing vials which enable the formulation of spin-coating inks with various compositions; a spin-coater for depositing inks on substrates to form thin films; an annealing station for variable-time annealing of thin films; a darkfield imaging station for imaging the thin films.

### *Materials*

Toluene (ACS grade) was purchased from Fisher Chemical, and was used without further purification. Acetonitrile (≥99.9%), 2-propanol (≥99.5%), acetone (≥99.5%), 4-*tert*-Butylpyridine (96%), Spiro-MeOTAD (99%), FK 102 Co(III) TFSI salt (98%, SKU 805203-5G), and Zinc di[bis(trifluoromethylsulfonyl)imide] ($Zn(TFSI)_2$, 95%) were purchased from Sigma Aldrich and were used without any futher purification. Extran 300 Detergent was purchased from Millipore Corporation. Titanium(IV) 2-ethylhexanoate (97%) was purchased from Alfa Aesar and was used without any further purification. White glass microscope slides (3" × 1" × 1 mm) were purchased from VWR International. Fused silica wafers (100 mm diameter, 500 µm thickness, double-side polished) were purchased from University Wafer.

Fused silica wafers and microscope slides were cleaned prior to thin film deposition. A solution of 1% *v/v* Extran 300 in deionized water was prepared. The substrates were sonicated successively in the diluted Extran 300, deionized water, acetone, and 2-propanol. Before each sonication step, the substrates were rinsed in the following solvent. Substrates were stored submersed in 2-propanol. Prior to use, the substrates were dried with filtered, compressed air and inspected by eye for defects.

*Organic thin film deposition*

Stock solutions of spiro-OMeTAD, FK102 Co(III) TFSI salt, Zn(TFSI)$_2$, and 4-*tert*-butylpyridine were prepared at 50 mg mL$^{-1}$ in 1:1 *v/v* acetonitrile/toluene. These stock solutions were combined using the robotic platform described above to form 150 µL of ink. 100 µL of ink was deposited by the robotic platform onto a microscope substrate rotating at 1000 rpm; rotation was maintained for 60 s following ink injection. The resulting thin films were then annealed for 0 to 250 s using a custom forced air annealer (an aluminum enclosure around heat gun, Model 750 MHT Products, Inc.). All of these procedures are described in more detail in ref.[27]

*Metal oxide thin film deposition*

Amorphous titanium oxide films were prepared by manual spincoating. The samples were prepared by pipetting 100µL of Titanium(IV) 2-ethylhexanoate solution (0.1 M, 2-propanol) onto cleaned fused silica wafers rotating at 3000 rpm; rotation was maintained for 30 s following ink injection. The resulting samples were irradiated with deep ultraviolet light (Atlantic Ultraviolet G18T5VH/U lamp – 5.8W 185/254 nm, ~2 cm from bulb, atmospheric conditions) for 15 minutes. After irradiation, the samples were transparent and highly refractive.

*Robotic darkfield imaging*

All darkfield images taken with the robot were captured with a FLIR Blackfly S USB3 (BFS-U3-120S4C-CS) camera using a Sony 12.00 MP CMOS sensor (IMX226) and an Edmund Optics 25mm C Series Fixed Focal Length Imaging Lens (#59-871). The C-mount lens was connected to the CS-mount camera using a Thorlabs CS- to C-Mount Extension Adapter, 1.00"-32 Threaded, 5 mm Length (CML05). The sample was illuminated from the direction of the camera using an AmScope LED-64-ZK ring light. For imaging, the lens was opened to f/1.4, and black flocking paper (Thorlabs BFP1) was placed 10 cm behind the sample.

*Bright-field microscopy*

All brightfield images were collected using an OLYMPUS LEXT OLS 3100 microscope operating in bright-field reflection mode using 5× and 20× objectives.

*Monotonic dewetting experiment*

To collect images of an organic thin film monotonically dewetting over time, a thin film of Spiro-OMeTAD and FK102 Co(III) TFSI salt was deposited (but not annealed) using the robotic platform as described above. A camera and lightsource were positioned above the sample in the same way as they were for the robotic darkfield imaging setup. A heat gun (Model 2363333, Wagner) was positioned to heat the sample from below at a 45° degree angle so as not to obscure the black background from the camera. To perform the experiment, the heat gun was turned on high and images were acquired every second for 100 seconds.

*Development of the DeepThin network*

The *DeepThin* CNN architecture (Fig. 1) was developed for the thin-film image analysis tasks described here and is inspired by the VGG16 CNN architecture[33]. Initially, *DeepThin* was trained using only one convolutional layer. The model complexity was iteratively increased until the model accuracy stopped improving.

The input layer to *DeepThin* is an image with 3 RGB color channels. *DeepThin* has several convolutional and pooling layers as detailed in Fig. 1. The first convolutional layer uses 32 filters with a 3×3×3 kernel to convolve over the image, creating an output of size 50×50×32. Zero padding is performed so that the resulting image size is identical to the input image size. The output of the convolutional layer is passed into a ReLU activation layer. This convolutional layer is repeated, as in the VGG16 model.

Next, a maximum pooling layer of kernel size 2×2 is convolved over the layer to generate a 25×25×32 output, returning the maximum value for a kernel. The two convolution layers and the pooling layer are repeated for a second time. The output of the second maximum pooling layer is flattened to a 2000×1 vector. This is followed by two fully connected layers of 20 neurons with ReLU activation functions, and a final layer that output defects classes by applying a sigmoid activation function. *DeepThin* is trained by minimizing an error function through back propagation using the stochastic gradient descent method. L2(Gaussian) and Dropout regularization was used to reduce interdependent learning amongst the neurons. Regularization reduces overfitting by adding a penalty to the loss function.

*DeepThin* was trained using the Adam optimizer[34], with an initial learning rate of 0.001 and a batch size of 100. Training loss and validation loss converged by 11 epochs.

## Data availability
The image datasets used in this study will be made permanently available on GitHub and are temporarily available for review purposes at [projectada.ca/data/imaging](projectada.ca/data/imaging). All other data supporting the findings of this study are available from the corresponding authors upon request.

## Acknowledgments
Brightfield microscopy was performed in the Centre for High-Throughput Phenogenomics at the University of British Columbia, a facility supported by the Canada Foundation for Innovation, British Columbia Knowledge Development Foundation, and the UBC Faculty of Dentistry. We thank Pierre Chapuis for exploratory work on image analysis and Gordon Ng for assistance with the dewetting experiment. For hardware and software contributions to the robotic platform, we acknowledge Michael Elliott, Ted Haley, Karry Ocean, Alex Proskurin, Michael Rooney, and Henry Situ.

## Author Contributions
C. P. B. supervised the project. N.T. developed the CNN. B. P. M., F. G. L. P., K. E. D. and T.D.M. used the robotic platform to create the 2600-image initial training set and to perform the doping and annealing experiments.  T. D. M. and K. E. D. labelled the datasets. F. G. L. P. performed the dewetting experiment. B. P. M. performed the microscopy. F. G. L. P. created the figures with input from N.T. and B.P.M. All authors contributed to the writing of the manuscript.

## Competing Interests statement
The authors declare no competing interests.

## Additional information
Correspondence and requests for materials should be addressed to C. P. B.

# Quantifying Defects in Thin Films using Machine Vision

N. Taherimakhsousi[1], B. P. MacLeod[1,2], F. G. L. Parlane[1,2], T. D. Morrissey[1,2], E. P. Booker[1], K. E. Dettelbach[1], C. P. Berlinguette[1-4]*

## Supplementary information

**Affiliations:**
[1]Department of Chemistry, The University of British Columbia, Vancouver, British Columbia, Canada
[2]Stewart Blusson Quantum Matter Institute, The University of British Columbia, Vancouver, British Columbia, Canada
[3]Department of Chemical & Biological Engineering, The University of British Columbia, Vancouver, British Columbia, Canada
[4]Canadian Institute for Advanced Research (CIFAR), MaRS Centre, Toronto, Ontario, Canada

*Email: cberling@chem.ubc.ca

**Supplementary figures**

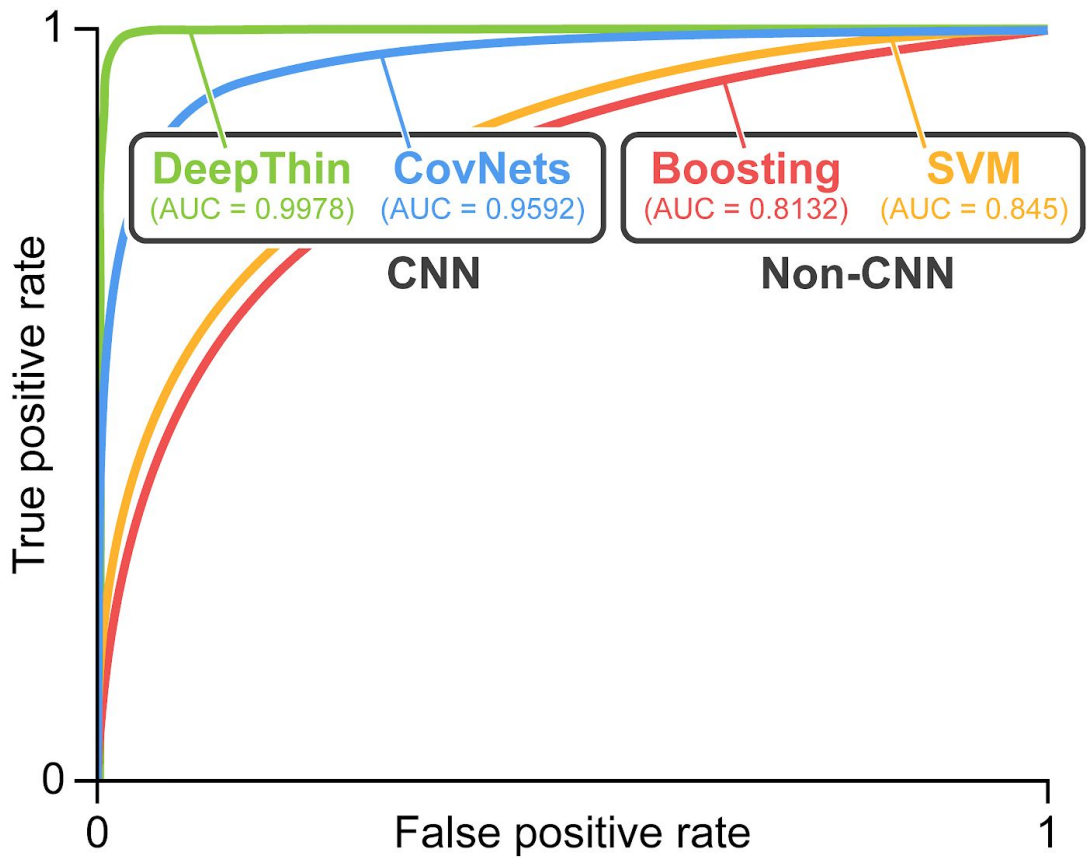

**Fig. S1 |** Receiver operating characteristic (ROC) curves for *DeepThin* compared to those for the road-crack detection algorithms reported by Zhang et. al[30].

**Supplementary tables**

**Table S1 |** Distribution of the darkfield images of organic films used for CNN training, validation and testing.

| Dataset | Cracks | Dewetting and No defects | Total | Dewetting | Cracks and No defects | Total |
|---|---|---|---|---|---|---|
| Training | 2468 | 7562 | 1030 | 5739 | 9032 | 14771 |
| Validation | 145 | 444 | 589 | 337 | 531 | 868 |
| Testing | 289 | 889 | 1178 | 674 | 1061 | 1735 |

**Table S2** | *DeepThin* performance on the training and testing portions of the darkfield organic film image set. TP: true positive, FN: false negative, TN: true negative, FP: false positive.

| Defect | Training Dataset | | | | | Testing Dataset | | | | | | | F1 score |
|---|---|---|---|---|---|---|---|---|---|---|---|---|---|
| | Accuracy | TP | FN | TN | FP | Accuracy | TP | FN | TN | FP | Precision | Recall | |
| Crack | 95% | 2095 | 373 | 7440 | 122 | 93.5% | 229 | 60 | 1124 | 17 | 0.93 | 0.79 | 0.86 |
| Dewetting | 94% | 1049 | 394 | 8568 | 464 | 94.4% | 626 | 48 | 1012 | 49 | 0.93 | 0.93 | 0.93 |

**Table S3** | Distribution of the bright-field microscopy images of organic films used for training, validation, and testing.

| Dataset | Particles | Cracks | Scratches | Dewetting | Total |
|---|---|---|---|---|---|
| Training | 2141 | 7229 | 2401 | 1817 | 13588 |
| Validation | 306 | 1032 | 259 | 972 | 2569 |
| Testing | 611 | 2065 | 518 | 1945 | 5139 |

**Table S4** | Description of the 20× magnification bright-field microscopy images of metal oxide films used for training, validation, and testing.

| Defect type | Samples | Train | Validate | Test |
|---|---|---|---|---|
| No Defects | 400 | 281 | 40 | 79 |
| Cracks | 578 | 405 | 58 | 115 |
| Particles | 898 | 629 | 90 | 179 |
| Non-uniform | 242 | 170 | 24 | 48 |
| | Total: 2118 | Accuracy: 87% | Accuracy: 87% | Accuracy: 84% |

**Table S5** | Descriptions of the 5× magnification bright-field microscopy images of metal oxide films used for training, validation, and testing.

| Defect type | Samples | Train | Validate | Test |
|---|---|---|---|---|
| No Defects | 1992 | 1395 | 199 | 398 |
| Particles | 2176 | 1524 | 217 | 435 |
| Non-uniform | 866 | 607 | 86 | 173 |
| Cracks | 362 | 254 | 36 | 72 |
| Scratches | 478 | 335 | 48 | 95 |
| | Total: 5874 | Accuracy: 90% | Accuracy: 88% | Accuracy: 86% |

**Table S6 |** *DeepThin* performance on classifying images from the testing portion of the bright-field organic microscopy image dataset reported as a confusion matrix.

|  |  | Actual |  |  |  |
|---|---|---|---|---|---|
|  |  | Cracks | Dewetting | Particles | Scratches |
| **Predicted** | Cracks | 1788 (86.5%) | 161 | 2 | 35 |
|  | Dewetting | 211 | 1693 (87.0%) | 12 | 38 |
|  | Particles | 15 | 17 | 597 (97.7%) | 0 |
|  | Scratches | 51 | 74 | 0 | 445 (85.9%) |

**Table S7 |** *DeepThin* performance on classifying images from the testing portion of the metal-oxide, 20× magnification, bright-field microscopy image dataset, reported as a confusion matrix.

|  |  | Actual |  |  |  |
|---|---|---|---|---|---|
|  |  | No defects | Particles | Non-uniform | Cracks |
| **Predicted** | No defects | 51 | 20 | 2 | 6 |
|  | Particles | 9 | 156 | 3 | 11 |
|  | Non-uniform | 0 | 7 | 41 | 0 |
|  | Cracks | 0 | 3 | 3 | 109 |

**Table S8 |** *DeepThin* performance on classifying images from the testing portion of the metal-oxide, 5× magnification, bright-field microscopy image dataset, reported as a confusion matrix.

|  |  | Actual |  |  |  |  |
|---|---|---|---|---|---|---|
|  |  | No defects | Particles | Non-uniform | Cracks | Scratches |
| **Predicted** | No defects | 381 | 17 | 0 | 0 | 0 |
|  | Particles | 17 | 403 | 15 | 0 | 0 |
|  | Non-uniform | 0 | 43 | 117 | 13 | 0 |
|  | Cracks | 1 | 11 | 18 | 41 | 0 |
|  | Scratches | 9 | 10 | 4 | 1 | 71 |

**Table S9 |** Performance comparison of different road-crack detection methods.

| Method | Precision | Recall | F1 score |
|---|---|---|---|
| SVM | 0.811 | 0.673 | 0.736 |
| Boosting | 0.736 | 0.759 | 0.747 |
| ConvNets | 0.869 | 0.925 | 0.897 |
| *DeepThin* (this work) | **0.992** | **0.987** | **0.991** |

**Table S10 |** Comparison between the architecture of *DeepThin* and the ConvNets road-crack detection neural network.

| Model | Conv1 (total) | Conv2 (total) | Conv3 (total) | Conv4 (total) | FC1 (total) | FC2 (total) | FC3 (total) |
|---|---|---|---|---|---|---|---|
| ConvNets | 4×4×3×48 +48 (2352) | 5×5×48×48 +48 (57648) | 3×3×48×48 +48 (20784) | 4×4×48×48 +48 (36912) | 3×3×48×200 +200 (86600) | 200×2 +2 (402) | N/A |
| *DeepThin* | 3×3×1×32 +32 (320) | 3×3×32×32+ 32 (9248) | 3×3×32×32+ 32 (9248) | 3×3×32×32+ 32 (9248) | 9×9×32×20 +20 (51860) | 20×20 +20 (420) | 20×2 +2 (42) |

CovNets: Kernel Size: 5×5, 4×4 and 3×3, Pool Size: 3×3 and 2×2, Total Params: 204696
*DeepThin*: All Kernel Size: 3×3, All Pool Size: 2×2, Total Params: 80386

**Table S11 |** Comparison of performance between *DeepThin* and previously reported algorithms for crack segmentation on images of cracked road surfaces from the Crack Forest Dataset.

| Method | Precision | Recall | F1 score |
|---|---|---|---|
| Canny[30] | 0.4377 | 0.7307 | 0.4570 |
| Local thresholding[31] | 0.7727 | 0.8274 | 0.7418 |
| CrackForest[32] | 0.7466 | **0.9514** | 0.8318 |
| Crack CNN[31] | 0.9119 | 0.9481 | 0.9244 |
| *DeepThin* (this work) | **0.9652** | 0.8915 | **0.9269** |

**Table S12 |** Comparison between the architecture of *DeepThin* and the CrackCNN road-crack segmentation neural network.

| Model | Conv1 (total) | Conv2 (total) | Conv3 (total) | Conv4 (total) | Conv5 (total) | FC1 (total) | FC2 (total) | FC3 (total) |
|---|---|---|---|---|---|---|---|---|
| Crack CNN | 3×3×3×16 +16 (448) | 3×3×16×16 +16 (2320) | 3×3×16×16 +16 (2320) | 3×3×16×32 +32 (4640) | 3×3×32×32 +32 (9248) | 7×7×32×64 +64 (100416) | 64×64 +64 (4160) | 64×25 +25 (1650) |
| *Deep Thin* | 3×3×1×32 +32 (320) | 3×3×32×32 +32 (9248) | 3×3×32×32 +32 (9248) | 3×3×32×32 +32 (9248) | N/A | 3×3×32×20+2 0 (5780) | 20×20 +20 (420) | 20×2 +2 (42) |

*CrackCNN*: All Kernel Size: 3×3, All Pool Size: 2×2, Total Params: 125202
*DeepThin*: All Kernel Size: 3×3, All Pool Size: 2×2, Total Params: 34306